\documentclass[11pt]{article}
\usepackage[latin1]{inputenc}
\usepackage{epsfig}
\usepackage{amsmath,amsfonts,latexsym, amssymb}

\newtheorem{satz}{Theorem}[section]
\newtheorem{defi}[satz]{Definition}

\newtheorem{koro}[satz]{Corollary}

\newtheorem{assumption}[satz]{Assumption}
\newtheorem{obdef}[satz]{Observation/Definition}
\newtheorem{conclusion}[satz]{Conclusion}
\newtheorem{ob}[satz]{Observation}

\newtheorem{propo}[satz]{Proposition}

\newcommand{\mcal}{\mathcal}
\newcommand{\mbf}{\mathbf}

\newcommand{\tit}{\textit}

\newcommand{\R}{\mathbb{R}}
\newcommand{\Z}{\mathbb{Z}}

\newcommand{\bewende}{$ \hfill \Box $}

\begin{document}
\thispagestyle{empty}
\begin{center}

{\LARGE{\bf Entanglement-Entropy for Groundstates,\\ Low-lying and 
Highly Excited Eigenstates\\ of General (Lattice) Hamiltonians}} 

\vskip 0.5cm

{\large {\bf Manfred Requardt }} 

\vskip 0.3 cm 

Institut f\"ur Theoretische Physik \\ 
Universit\"at G\"ottingen \\ 
Friedrich-Hund-Platz 1 \\ 
37077 G\"ottingen \quad Germany\\
(E-mail: requardt@theorie.physik.uni-goettingen.de)

\end{center}

\vspace{0.3 cm}
\noindent Short title: Entanglement-Entropy\\[0.2cm]
 
\begin{abstract} {\small We investigate the behavior of
    entanglement-entropy on a broad scale, that is, a large class of
    systems, Hamiltonians and states describing the interaction of
    many degrees of freedom. It is one of our aims to show which
    general characteristics are responsible for the different types of
    quantitative behavior of entantglement-entropy. Our main lesson is
    that what really matters is the degree of degeneracy of the
    spectrum of certain nearby reference Hamiltonians. For
    calculational convenience we study primarily systems defined on
    large but finite regions of regular lattices. We show that general
    vector states, being not related to some short-range Hamiltonian
    do not lead in the generic case to an area-like behavior of
    entanglement-entropy.  The situation changes if eigenstates of a
    Hamiltonian with short-range interactions are studied. We found
    three broad classes of eigenstates. Global groundstates typically
    lead to entanglement-entropies of subvolumes proportional to the
    area of the dividing surface. Macroscopically excited
    (vector)states have in the generic case an entanglement-entropy
    which is proportional to the enclosed subvolume and, furthermore,
    display a certain Gibbsian behavior. Low-lying excited states, on
    the other hand, lead to an entanglement-entropy which is usually
    proportional to the logarithm of the enclosed subvolume times the
    area of the dividing surface . Our analysis is mainly based on a
    combination of concepts taken from the perturbation theory of
    Hamiltonians and certain insights coming from the foundations of
    quantum statistical mechanics.  }
\end{abstract} 

\newpage

\setcounter{page}{1}
\section{Introduction}
The microscopic origin of entropy in, for example, black hole physics
and the so-called \tit{area law} is still to some extent kind of a
mystery. Or more precisely, there exist a variety of different
explanations. For a nice discussion in form of a Galilean trialogue
see \cite{Jacobson}. Another very readable overview is given in
\cite{Wald}, a collection of theses can also be found in
\cite{Sorkin}. Of perhaps even more interest is the proposed
\tit{holographic principle} (and the range of its validity), which
discusses the emergence of entropy on a very fundamental level of
physics, that is, ``empty'' space (-time) and its \tit{vacuum
  fluctuations}. We do not attempt to give a complete list of
references (as we focus in the following on a related but slightly
different question), cf. for example the nice review \cite{Bousso} and
the many references therein.

What is at least clear is that the black hole horizon divides
space-time into \tit{exterior} and \tit{interior} regions. Hence, one
might venture the idea that a basic role in this question should be
played by some form of \tit{entanglement} of two (in some macroscopic
respect) separated regions. Such a form of entanglement can however be
realized on different levels of our theoretical description, ranging
from more ordinary ones (standard quantum mechanics or quantum field
theory) to more pristine ones ( e.g. the notorious Planck scale).  On
this point people widely disagree at the moment. As to our personal
point of view, we think that the holographic principle indicates a
drastic change of the statistical mechanical preassumptions usually
taken for granted on the more ordinary scales as e.g. \tit{locality},
good clustering conditions or sufficient decay of correlations etc.,
if one enters the fundamental regime of the Planck scale. Some very
sketchy remarks can be found at the end of \cite{Requ}, a more
detailed analysis is forthcoming (\cite{BH}). Be that as it may, one should
emphasize the following:
\begin{ob}The following two situations are different from a logical
  point of view, for one, entanglement-entropy induced by a fixed
  global state as e.g. the groundstate of a given Hamiltonian, for
  another, the maximally possible quantum of entropy or information,
  $S_{max}$, which can be stored in a given volume, the question which
  is adressed in the holographic principle. In the latter case also
  the highly excited states of a Hamiltonian should be investigated, a
  problem we explicitly address in the following sections.
\end{ob}

In any case, in a first step, it seems to be useful to investigate
ordinary models of quantum (field) theory and see to what extent their
entropic behavior reflects some of the properties condensed in notions
like area law and/or holographic principle. A certain kind of
area-law-like behavior shows up already in ordinary quantum field
theory, see e.g. \cite{Brustein1} or \cite{Brustein2}. Similar
phenomena were also observed in \cite{Requ1}, \cite{Requ2} and more
recently in \cite{Requ}. As this is an interesting point, we will
treat it in more detail elsewhere.

To make the problem more accessible, we start with the investigation
of systems of many degrees of freedom living on a large regular
lattice (of arbitrarily small but finite lattice constant). In
\cite{Bombelli} or \cite{Srednicki} or the more recent \cite{Plenio1}
large arrays of coupled harmonic oscillators and their respective
groundstates have been studied, with some sort of continuum limit
performed in the end, yielding the vacuum state of e.g. a Klein-Gordon
field theory. Note however that taking a continuum limit is quite
delicate since, in order to get a finite result, some form of
\tit{cut-off} is necessary (as to other possible interesting
applications to quantum field theory see for example \cite{Schroer}).

As everybody knows, the harmonic oscillator is an extremely
well-behaved quantum system. The same holds for regular arrays of
coupled harmonic oscillators. There exists a large arsenal of methods
to extract useful information from these model systems. But
nevertheless, the question of \tit{entanglement-entropy} of e.g. the
groundstate of this well-behaved model system turns already out to be
quite intricate. This may, among other things, have its roots in the
observation that the concept of entanglement-entropy and its
quantitative behavior is not so easy to visualize, one reason being,
that it is not a truely local concept.

It was found that if one divides a large volume, $V$, into $V_1\cup
V_2$ and restricts the groundstate, $\Psi_0$, over $V$ to the
subvolumes $V_1,V_2$, the corresponding \tit{density matrices},
$W_1,W_2$, have an entropy, $S=-\sum w_i\ln w_i$, which is
proportional to the area of the dividing surface between $V_1$ and
$V_2$! One may therefore speculate that this observation has something
to do with the area law in black hole physics and corresponding
arguments were advanced in the two above cited papers. 

If one persues such an idea, various questions immediately suggest
themselves.\\
i) In what respect is the groundstate of the Hamiltonian of a large
system (i.e., many degrees of freedom being involved) a particular
state in this special context. Put differently, what particular
property of a global vectorstate (pure state) influences the entropic
behavior of the \tit{partial traces},
$W_1,W_2$? \\
ii) In what respect does the picture change if we go over to
\tit{lowly} or \tit{highly excited} eigenstates of such a
Hamiltonian? \\
iii) Is the result, found for an array of harmonic oscillators,
generic, that is, is it to a high degree independent of the particular
model system being studied and what are the really
important prerequisites?\\
These are the questions we will adress in the following sections. A
preliminary qualitative answer may perhaps already be given. Perhaps a
little bit surprisingly, what really seems to matter is the degree of
degeneracy of the spectral values of certain nearby reference
Hamiltonians. For groundstates, for example, this means that the
area-behavior is a result of their usual \tit{non-degeneracy}. 

We begin our investigation with defining the general framework of
systems, living on large regular lattices. We then discuss in section
3 several general properties of the notion of
\tit{entanglement-entropy}. We mention for example that a Gibbs state
over $V_1$ can always be extended to a pure vector state, $\Psi$, on
$V$ by enlarging the system on $V_1$ with $V=V_1\cup V_2$.  That is,
the Gibbs state is now the partial trace of a global vectorstate. As,
typically, such Gibbs states have an entropy which is proportional to
the volume, $V_1$, this shows that, in general, entanglement-entropy
can \tit{not} be proportional to the area of the surface, dividing
$V_1$ and $V_2$.  We learn from this simple observation that, in order
to get such a special result, we do need some more specifications.

This observation teaches us yet another lesson. One sometimes hears
the argument that entanglement-entropy has to be proportional to the
area of the dividing surface as this is the only quantity which is the
same for both subvolumes, $V_1$ and $V_2$ while the volumes themselves
can of course differ from each other (note that, perhaps a little bit
surprisingly, entanglement-entropy is always the same for $V_1$ and
$V_2$). This argument is obviously incorrect. What in general may
happen is that we have something like the following
\begin{equation}\label{volume}S(W_1)=\alpha_1\cdot |V_1|=\alpha_2\cdot |V_2|=S(W_2)
\end{equation}
with $W_1,W_2$ the density matrices representing the total state,
$\Psi$, on $V_1,V_2$ respectively. That is, both entropies may happen
to depend linearly on the volume but with certain (volume-dependent)
adjusting prefactors. As long as the volumes are finite, there is no
unique answer as to the geometric dependencies. We need some kind of
\tit{thermodynamic limit}. In \cite{Plenio1} one kind of such limit is
discussed. One can, for example, keep the dividing surface fixed and
let the subvolumes approach infinity in the orthogonal direction, or,
on the other hand, it may become necessary, to scale them in a
prescribed way with the prefactors remaining basically constant. In
the former case we encounter a linear dependence on the area of the
dividing surface if the limit is finite, in the latter (scaling) case we may
get a linear asymptotic dependence of $S(W_i)$ on the volume.

At the end of that section we introduce a certain \tit{coarse-grained}
version or approximation of the notion of entanglement-entropy for the
subvolumes $V_1,V_2$, which, in some sense, resembles kind of a
microcanonical ensemble version of entropy. In our particular case it
is related to the logarithm of the dimension of certain Hilbert
subspaces onto which the pure vectorstate happens to be essentially
projected when calculating the relative trace. This is an important
simplification: While the original standard notion of entropy,
$S:=-\sum w_i\cdot \ln w_i$, is deceptively simple at first glance,
this is a very missleading impression. Even in the few cases where the
Boltzmann weights, $w_i$, are known, what is usually more important is
their (frequently huge) degeneracy and/or the local density of states,
which in general are not at our disposal.\\[0.3cm]
Remark:We recently learned that this approximation scheme we will
discuss below has a certain resemblance to an approximation method
used in the \tit{density matrix renormalisation group, DMRG} and which
can roughly be described as finding the \tit{essential subspace}. It
is also of relevance in data compressing and the like (see e.g.
\cite{White},\cite{Nielsen},\cite{Legeza},\cite{Farkas}).\vspace{0.3cm}

While we attempt to treat the problem of entanglement-entropy on quite
a broad scale (as far as space dimension and class of hamiltonians are
concerned), there are interesting topics we do not! address. In our
scenario we deal with a \tit{spatial} tensor product structure of the
underlying Hilbert space. Most notably in condensed matter physics
there exist applications where other kinds of configurations
prevail. One example is the case of fermionic quantum particles where
the \tit{antisymmetrization postulate} prevents such a
representation. In that situation one can for example work in a
\tit{mode representation}; (basically \tit{Fock space}) and discuss
entanglement in such a mode space. Some work in this direction may be
found in the review article \cite{Fazio}, \cite{Zanardi}, \cite{Shi}
or \cite{Wolf}. We think that parts of our strategy can also be applied
in this field of research but this is not done in the present paper. 

Recently, Shi (\cite{Shi2}) attempted to extend such a mode-version of
entanglement to relativistic quantum field theory. This is certainly
an interesting programme but we would like to hint at a number of
impending difficulties and obstacles. In free field theories and
Fourier space, occupation number representations for example have a
well-defined meaning. In coordinate space and, a fortiori, interacting
theories such an approach is met with great difficulties due to the
intricacies of the relativistic quantum regime. For one, expressions
like $\phi^{\dagger}(x)\phi(x)$ are no longer operators but
operator-valued distributions. The same applies to eigenstates,
``localized at $x$''. What one can use instead are \tit{localized
  algebras} of observables living in regions $\mcal{O}$ of space-time.
One may then restrict e.g. the vacuum state to such local algebras,
thus getting a \tit{mixed state}.

It is however obvious that in general a naive splitting into, say, a
bipartite system, having kind of a tensor-product structure, does not
work (due to quantum non-locality and other problems; a catchword
being the \tit{Reeh-Schlieder Property}). In any case, the mathematics
becomes extremely involved as most of the naive expressions become
infinite (for example \tit{traces}). Only for a few scenarios one can
say more (like e.g. the \tit{Rindler-wedge}). The interested reader
may consult section V,5 of the book by Haag (\cite{Haag}),\cite{Schroer} or
\cite{Wald} for a general orientation.

Furthermore, our arguments rely to a certain extent (at least in the
form presented here) on short-range correlation and short-range
interactions. This is the field where one may expect that results are
\tit{generic} to some extent. Therefore our arguments do not
(immediately) apply to \tit{quantum critical systems}, i.e.  systems
displaying \tit{quantum phase transitions} at temperature zero. These are
presently mostly studied in low dimensions (space dimension frequently
being one) by employing usually quite model-specific techniques. An
incomplete selection of papers is
\cite{Cardy1},\cite{Cardy2},\cite{Peschel}, \cite
{Vidal},\cite{Korepin1},\cite{Korepin2},\cite{Korepin3}.

Given the generality of our admitted class of models, it is not our
aim and out of the question to really compute in every case the
complete set of eigenvalues and their degeneracy or spectral density.
This holds the more so for the v.Neumann entropy. We rather view the
merit of our approach in providing a general scheme which only relies
on a few characteristics of the respective models. As a consequence,
our arguments and estimates do not yield precise numerical results but
typically make statements about entanglement-entropy in some leading
order (for more details see below).

For the same reason (a large class of different admissible
Hamiltonians) we develop in section \ref{perturbation} in quite some
detail various concepts belonging to the perturbational analysis of
Hamiltonians. This turns out to be a very intricate subject matter in
our context. The main reason is that, due to the dense distribution of
eigenvalues typical for Hamiltonians of a large number of degrees of
freedom, ordinary perturbation theory is only of a limited value.

In the last three sections we then deal with our main topic in a more
quantitative way. In section \ref{groundstate} we treat the case of
the groundstate of a Hamiltonian with \tit{finite range} interactions.
In the following section we analyse the situation for eigenstates
which are highly excited, that is, for energies which are a
macroscopic distance apart from the groundstate energy. In the last
section we study, on the other hand, eigenvalues of the Hamiltonian
which lie in the vicinity of the groundstate energy. We find the
following:\\
i) Groundstates come with an entanglement-entropy which is
proportional to the area of the dividing surface.\\
ii) Highly (i.e. macroscopically) excited states lead in the
\tit{generic} case to entanglement-entropies which are proportional to
the volume of the system. The notion \tit{generic} is explained in the
respective section. Furthermore, the restriction of the pure
vectorstate to the respective subvolumes share a couple of properties
with ordinary temperature states. As for
more detailed qualifications see the corresponding section.\\
iii) Low-lying excited eigenstates states have an entanglement-entropy
which is roughly proportional to the logarithm of the corresponding
volume times the area of the dividing surface.

To sum up what we think have been important ingredients in our
analysis, the arguments strongly rely on the finite range of the
interactions, that is some version of \tit{locality}. Furthermore, we
assume the system to be away from a \tit{(quantum-)critical state}
where local degrees of freedom are typically long-range correlated. In
this special regime results happen to be somewhat model dependent and
our more general arguments need not automatically apply. As to the
applied methods, two tools stand out. First, a combination of various
perturbational arguments, second, a couple of technical ideas which
have proved their worth already in the foundations of \tit{statistical
  mechanics}.

We want to add a further remark. One might perhaps get the idea that
high entanglement is necessarily related to long-range correlations.
That this is not the case can be inferred from our following
investigations. It is for example easy to construct a highly entangled
eigenstate of a short-range Hamiltonian. What is only important is
that the respective eigenvalue happens to be highly degenerate or that
the full Hamiltonian is the perturbation of such a Hamiltonian with
highly degenerate eigenstates. This is exploited in section
\ref{highly}. Note that, typically, most eigenvalues of Hamiltonians
of many degrees of freedom are highly degenerate, an exception being
usually the ground state. For general vector states there is even less
correlation between entanglement and range of correlations.  An
interesting observation in this direction is for example also made in
\cite{Verstraete1} and \cite{Verstraete2} where these two notions are
compared with each other.

A related line of ideas is pursued in \cite{Shi3}. Again Shi
concentrates mainly on mode expansions of many body systems and
studies in particular the regime of phase transitions and spontaneous
symmetry breaking. For example, in the case of Bose-Einstein
condensation he remarks that a macroscopic occupation of the ground
state leads to what he calls \tit{disentanglement} with the rest of
the modes. This is reasonable as the ground state occupation number
operator becomes almost a classical observable in this situation. He
further addresses the relation between entanglement and range of
correlations in this context. It conforms with our observations that
long-range correlations make it in general difficult to divide a
system into, for example, a weakly correlated bipartite system, while
high entanglement does not necessarily imply long-range correlations.

\section{Some General Framework and Estimates}
Concerning this section a standard reference is for example
\cite{Ruelle}. To avoid irrelevant complications it is reasonable to
choose as the general context discrete dynamical models living on a
regular lattice, $\Z_a^d$, with lattice constant $a$ and space
dimension $d$. The points of the lattice are denoted by $x_i$, the
Hilbert space sitting at $x_i$ by $\mcal{H}_i$. $\mcal{H}_i$ may be
finite or infinite dimensional (as is for example the case for an
array of harmonic oscillators). If all the $\mcal{H}_i$ have the same
finite dimension it is denoted by $D$.

In the following we are interested in large but finite subsets of
$\Z_a^d$, denoted by $V$. For convenience $V$ may be chosen as a
$d$-dimensional cube of side length $L$, containing $N_V=(L/a)^d$
lattice points. The general case of regions of arbitrary shape can of
course be treated in more or less the same way apart from some
(irrelevant) numerical and technical details. The Hilbert space,
$\mcal{H}_V$, over $V$ is the tensor product
\begin{equation}\mcal{H}_V=\bigotimes_{x_i\in V}\mcal{H}_i    \end{equation}
spanned by tensor-monomials
\begin{equation}v=v_1\otimes\cdots\otimes v_{N_V}\;,\;v_i\in
  \mcal{H}_i \end{equation}
 If one chooses a basis, $\{e_i^{l_i}\}$,
in each $\mcal{H}_i$, general vectors in $\mcal{H}_V$ are linear
combinations of the basic vectors
\begin{equation}e_1^{l_1}\otimes\cdots\otimes
  e_{N_V}^{l_{N_V}}\;,\;1\leq l_i\leq D        \end{equation}
(in case the dimension of all $\mcal{H}_i$ is finite and equal). Then the
dimension of $\mcal{H}_V$ is $D^{N_V}$. The scalar product on
$\mcal{H}_V$ is given by
\begin{equation}(v_1\otimes\cdots\otimes
  v_{N_V}|v'_1\otimes\cdots\otimes v'_{N_V})  :=\prod_{x_i}(v_i|v'_i)     \end{equation}
and (multi)linearly extended.

Hamiltonians are now defined as follows. If $A_i$ are operators (or matrices)
on $\mcal{H}_i$, operator monomials on $\mcal{H}_V$ are given by 
\begin{equation}\bigotimes_{x_i}A_i\,:\bigotimes v_i\mapsto \bigotimes
  A_i\circ v_i\end{equation}
and (multi)linearly extended on more general vectors. Refraining from
a too abstract approach we restrict ourselves to Hamiltonians given by
a sum over \tit{one-point, two-point,\ldots,
  k-point-interactions}. This means
\begin{equation}\sum_{x_i}\Phi^1(x_i)\,,\,\sum_{x_i,x_j}\Phi^2(x_i,x_j)\,,\,\ldots        \end{equation}
where 
\begin{equation}\Phi^1(x_i)=A_i^{(1)}\;,\,\Phi^2(x_i,x_j)=A_i^{(2)}\otimes
A_j^{(2)}\;,\;\ldots\end{equation}
\begin{assumption}To avoid unneccessary complications we assume that
  only interaction terms up to a certain finite order occur in the
  Hamiltonian and that all the occurring operators $A_i$ are hermitean.
\end{assumption}

As all the $A_i,A_j$ do commute for $i\neq j$, the interaction terms
defined above are hermitean as is their sum.  Typical examples are
spin systems. One of the simplest examples is 
\begin{equation}H=J\cdot\sum_i \overrightarrow{\sigma}_i\otimes
  \overrightarrow{\sigma}_{(i+1)}=:J\cdot\sum_i
  \overrightarrow{\sigma}_i\overrightarrow{\sigma}_{(i+1)}
\end{equation} with $\sigma$ denoting the usual Pauli-matrices and
extension to higher dimensions being straightforward.

Usually one makes the additional assumption that the interaction has a
\tit{finite range} and is \tit{translation invariant}.
\begin{defi}The interaction is called translation invariant if with
 \\ $\Phi^k(x_{i_1},\ldots,x_{i_k})$ also
  $\Phi^k(x_{i_1}+r,\ldots,x_{i_k}+r)$ occurs in the Hamiltonian with
  $r$ some lattice vector and both being the same operators.
\end{defi}
\begin{defi}The interaction is called to have a finite range,
  $\delta$, if with one lattice point fixed, e.g. $x_{i_1}$, only
  finitely many members $\Phi^k(x_{i_1},y_{i_2},\ldots,y_{i_k})$
  different from zero exist in $H$ when the $y_{i_l}$ vary over
  $\Z_a^d$, all $y_j$ having a distance from $x_i$ which is
  $\leq\delta$.
\end{defi}
We not that we give these details only for consistency reasons,
whereas only the general properties of such Hamiltonians will be of
relevance in the following like their translation invariance and the
finite range.

What will be of interest for the following discussion is the
restriction of a general Hamiltonian, $H$, to a certain subvolume
$V\subset \Z_a^d$ or the corresponding sub-Hilbert-space $\mcal{H}_V$.
\begin{defi}The restriction of a general Hamiltonian, $H$, of the kind
  described above to a subvolume $V\subset \Z_a^d$ or $\mcal{H}_V$ is
  the operator $H_V$ which consists of all the interaction terms which
  lie in $V$, i.e., only $k$-tuples $(x_{i_1},\ldots,x_{i_k})$ are
  admitted with all the $x_{i_l}\in V$.
\begin{equation}H_V:=\sum_{k;x_{i_l}\in V}\Phi^k(x_{i_1},\ldots,x_{i_k})=:\sum_{n(V)}\Phi^k    \end{equation}
($n(V)$, the number of terms in the sum being of order $O(|V|)$).
\end{defi}
Another important part of the total $H$ relative to a given volume
$V$ is the \tit{boundary contribution} $H_V^{bd}$.
\begin{defi}By $H_V^{bd}$ we denote the part of the total $H$ which
  consists of all interaction terms which have both lattice points in
  $V$ and the dual set $V':=\Z_a^d\setminus V$.
\begin{equation}H_V^{bd}:=\sum_{k;x\in V,y\in V'}\Phi^k(x_{i_1},\ldots;y_{i_l},\ldots,y_{i_k})     \end{equation}
\end{defi}
\begin{ob}It will be important in the following that with the
  Hamiltonian having finite range and $k\leq K$, the number of terms
  in $H_V^{bd}$ is of order (area of boundary of $V$). We denote this
  number of terms by $n(V,V')$.
\end{ob}
We can make more detailed statements if we concentrate on the large
subclass of models with all the $\Phi^k$ bounded operators (which
holds for example if all the $\mcal{H}_i$ are finite dimensional).
This together with the assumed translation invariance and the above
observation allow us to make the following important operator
estimates.
\begin{propo}Under the preceding assumptions we have
\begin{equation}\parallel H_V\parallel\leq C\cdot n(V)\;,\; \parallel
  H^{bd}_V\parallel\leq C'\cdot n(V,V')    \end{equation}
where in the generic case, which we usually assume to prevail, both
norms are actually proportional to the volume, the area of the boundary,
respectively.\\
In case the local Hilbert spaces $\mcal{H}_i$ are finite dimensional, all the
occurring Hamiltonians are large but nevertheless finite
hermitean matrices for finite $V$, hence having a discrete spectrum,
with the number of eigenvalues (counting multiplicity) being the same
as the dimension of the respective Hilbert spaces (the eigenvalues
being the zeros of the characteristic polynomial).
\end{propo}
It is obvious that this latter observation greatly simplifies the
quantitative analysis.

We will in the following mainly discuss the case where the local
Hilbert spaces have finite dimension. As the extension of the results
to the case where the local Hilbert spaces, $\mcal{H}_i$ have infinite
dimension is not entirely straightforward (like e.g. arrays of coupled
harmonic oscillators), we make some corresponding remarks in appendix
\ref{infinite}.

\section{Some Properties of the Partial Trace and Entanglement
  Entropy}
Let us take a vector state, $\Phi$, from the Hilbert space
$\mcal{H}_V$ with $V$ large (or macroscopic). Let us divide $V$ into
$V_1\cup V_2$ with the respective Hilbert spaces,
$\mcal{H}_1,\mcal{H}_2$. We have
\begin{equation}\mcal{H}_V=
  \mcal{H}_1\otimes\mcal{H}_2\;,\;\mcal{A}_V=\mcal{A}_1\otimes\mcal{A}_2   \end{equation}

We can evaluate the pure state, $\Phi$ on the restrictions
$\mcal{H}_1,\mcal{H}_2$:
\begin{equation}\omega_1(A^{(1)}):=(\Phi|(A^{(1)}\otimes\mbf{1})\Phi)\;,\;
\omega_2(B^{(2)}):= (\Phi|(\mbf{1}\otimes B^{(2)})\Phi)  \end{equation}
which defines the states $\omega_1\,,\,\omega_2$ on
$\mcal{H}_1,\mcal{H}_2$. Choosing a basis $e_i\otimes e_j'$ in the
tensor product $\mcal{H}_V=\mcal{H}_1\otimes\mcal{H}_2$ we get
\begin{equation}\Phi=\sum c_{ij}\,e_i\otimes e_j'  \end{equation}
and
\begin{equation}
(\Phi|(A\otimes\mbf{1})\Phi)=\sum_{m,i,j}\overline{c}_{im}c_{jm}(e_i|A
  e_j)=:\sum_{i,j}b_{ji}(e_i|A e_j)  \end{equation}
with
\begin{equation} b_{ji}=\sum_{m=1}^{dim(\mcal{H}_2)}c_{jm}\overline{c}_{im}
 \end{equation}
This can be rewritten as
\begin{equation}\label{18}\omega_1(A)=Tr(W_1\cdot A)\quad\text{with}\quad W_1:=\sum_{ij}b_{ji}|e_j><e_i|    \end{equation}

The following result is remarkable and leads to the definition of the
notion of \tit{entanglement-entropy}.
\begin{satz}\label{trace}The partial traces $W_1\,,\,W_2$ have the
  same spectrum
  and the eigenvalues $\neq 0$ have the same multiplicity while the
  respective zero-eigenvalues may have different degeneracies,
  depending on the in general different dimensions of
  $\mcal{H}_1\,,\,\mcal{H}_2$.
\end{satz}
\begin{koro}The (entanglement-)entropies of the states
  $\omega_1\,,\,\omega_2$ or $W_1\,,\,W_2$ are the same, i.e.
  \begin{equation}S_2(W_2)=S_1(W_1):=-\sum \lambda_i\cdot \ln\lambda_i
  \end{equation}
 while the respective volumes or dimension of Hilbert spaces can be
 very different.
\end{koro}
As the full proof of this important result is frequently omitted in the
literature or is incomplete we give, for the convenience of the
reader, our own version in appendix \ref{entanglement}.

On the other hand we know that the usual thermodynamic entropy is an
extensive quantity and depends in general linearly on the volume. So
let us assume we have a Gibbs-state on the volume $V_1$ with
Hamiltonian $H_1$, its eigenvalues and eigenstates being
$E_i\,,\,\psi_i$. That means:
\begin{equation}\omega_1(A):=Tr(e^{-\beta H_1}\cdot A)/Tr(e^{-\beta H_1})   \end{equation}
with $\beta$ the inverse temperature $1/kT$. It follows
\begin{equation}S_1=-\sum p_i\ln p_i\;,\;p_i=e^{-\beta E_i}/\sum e^{-\beta E_i}    \end{equation}
and  
\begin{equation}S_1(V_1)\sim V_1  \end{equation}

We now adjoin a volume $V_2$ with a Hilbert space $\mcal{H}_2$ of
sufficiently high dimension. We define the following vector, $\Psi$,
in $\mcal{H}_1\otimes\mcal{H}_2$.
\begin{equation}\Psi:=\sum \sqrt{p_i}\,\psi_i\otimes e_i  \end{equation}
($e_i$ spanning a basis in $\mcal{H}_2$ and degeneracies being
included). We have
\begin{equation}(\Psi|A^{(1)} \Psi)=Tr(e^{-\beta H_1}\cdot A^{(1)})/Tr(e^{-\beta H_1})  \end{equation}
We conclude:
\begin{ob}Every Gibbs state over $\mcal{A}_1$ can be represented by a
  vector state in a sufficiently large ambient Hilbert space.
  Restricted to $\mcal{H}_1$ this vector state is the partial trace,
  but we see from our above reasoning that the entanglement-entropy of
  $\Psi$ relative to $V_1$ is now proportional to the volume $V_1$.
\end{ob}

It is important for the physical understanding and intuition to get a
better feeling how entanglement-entropy is affected in both a
quantitative and qualitative way by different physical conditions. It
seems to be useful to introduce concepts which are, at least in a
rough sense, of a similar character as entanglement-entropy because the
latter can only be calculated in very few special cases. We think in
this context of the relation between, for example, the canonical and
the microcanonical ensemble in statistical mechanics.

For one, it is obvious that the entropy of a state, $\Psi_0$, reduced
to, say, $V_1$ or $V_2$, does not depend on the choice of Hilbert
space bases in the corresponding Hilbert spaces. So, in the following,
we will frequently subdivide $V_1$ or $V_2$ further into, say,
$V_1=V_1'\cup V_1''$ with $V_1''$ a boundary layer in $V_1$
neighboring upon the interface, separating $V_1$ and $V_2$. One can
then equally well choose a basis in $\mcal{H}_1$ by forming the tensor
product of the respective bases in $\mcal{H}'_1$ and $\mcal{H}''_1$
etc.

For another, a famous theorem of E.Schmidt and von Neumann
(\cite{Schmidt},\cite{Neumann}) states that $\Psi_0$ in
$\mcal{H}_1\otimes\mcal{H}_2$ can be represented in the special form
\begin{equation}\Psi_0=\sum_i\sqrt{\lambda_i}\cdot\phi_{i,1}\otimes\phi_{i,2}
\end{equation}
with $\phi_{i,1}\,,\,\phi_{i,2}$ particular orthonormal bases in
$\mcal{H}_1\,,\,\mcal{H}_2$ (the so-called Schmidt-basis; this was
also exploited in \cite{Ekert}).\\[0.3cm]
Remark: In modern parlance this is nothing but the theorem that a
compact operator can be put into such a canonical spectral form. Note
in this context that $\Psi_0$, viewed as an operator from $\mcal{H}_2$
to $\mcal{H}_1$, belongs to the \tit{Hilbert-Schmidt-class} as a
consequence of its normalisation as a vector.\\[0.3cm]
$\Psi_0$ reduced to $\mcal{H}_1$ then yields:
\begin{equation}(\Psi_0|A^{(1)}\Psi_0)=\sum_i\lambda_i\cdot
  (\phi_{i,1}|A\phi_{i,1})=Tr(W_1\cdot A)      \end{equation}

What is important in this particular representation is that both
systems of vectors are orthonormal.  With
$\Phi=\sum_{ij}c_{ij}e_i\otimes e_j'$ we can of course always write
\begin{equation}\label{1}\Phi=\sum_i e_i\otimes \phi_i= \sum_j \psi_j\otimes e'_j   \end{equation}
with
\begin{equation}\label{2}\phi_i=\sum_j c_{ij}\,e'_j\;,\;\psi_j=\sum_i c_{ij}\,e_i     \end{equation}
or
\begin{equation}\Phi=\sum_i\parallel \phi_i\parallel\!\cdot
  e_i\otimes\hat{\phi_i}=\sum_j\parallel\psi_j\parallel\!\cdot
  \hat{\psi_j}\otimes e_j'     \end{equation}
$\hat{\phi_i},\hat{\psi_j}$ being the corresponding unit vectors.

The reduction to $\mcal{H}_1\,,\,\mcal{H}_2$ then yields:
\begin{equation}(\Phi|A^{(1)}\Phi)=\sum_j(\psi_j|A\psi_j)\;,\;(\Phi|B^{(2)}\Phi)=\sum_i(\phi_i|B\phi_i)      \end{equation}
or
\begin{equation}(\Phi|A^{(1)}\Phi)=\sum_j\parallel\psi_j\parallel^2\cdot (\hat{\psi_j}|A\hat{\psi_j})\;,\;(\Phi|B^{(2)}\Phi)=\sum_i\parallel\phi_i\parallel^2\cdot(\hat{\phi_i}|B\hat{\phi_i})      \end{equation}
Remark:Note that the state, given as an expectation functional, is of
course identical to the state in e.g. formula (\ref{18}). But now the
way the state is concentrated in a subspace (either spanned by the
$\hat{\psi_j}$ or the $\hat{\phi_i}$) of the total space is made transparent.  
Note that the subspace may depend on the chosen basis in contrast to
its dimension (see below)\vspace{0.3cm}

Trying to relate the canonical version of the notion of entropy,
$S(W):=-\sum_i p_i\ln p_i$ with some other (perhaps coarser) concept,
we were inspired by the beautiful analysis of the entropy concept, as
it is laid out in \cite{Landau}, sect.7 of chapt.1. It is shown there
that the natural quantity which is relevant in this context is the
number of microscopic quantum states, $\triangle\Gamma$, a macrostate
is smeared over or, in other words, the number of microstates which
essentially contribute in a macrostate (taken with equal weights). It
is then shown in \cite{Landau} that in the regime of equilibrium
statistical mechanics, the logarithm of this quantity coincides with
the canonical notion of entropy given above, but this result is far
from trivial. We see that $\ln\triangle\Gamma$, giving equal apriori
weight to the members of a certain selected sample of quantum states,
implements the philosophy of the microcanonical ensemble picture.

In the following sections we are primarily interested in
leading-order-estimates, that is, estimates of quantities on a scale
given by macroscopic volumes or by the area of some bounding surface
etc. So, by inspecting the above formulas, we see that a rough notion
which reflects the number of ``different'' states being involved in
the reduction of a vector state to a partial trace over a certain
subvolume, is given by the dimension of the respective Hilbert
subspace the contributing vectors are lying in. This however needs
more qualifications.

In a first step we regard the above representation (formulas
(\ref{1}),(\ref{2}) of the vector $\Phi$ as a map from the Hilbert
spaces $\mcal{H}_1\,,\,\mcal{H}_2$ to the corresponding subspaces
spanned by the (in general non-orthogonal) $\phi_i\,,\,\psi_j$. These
maps (more properly, the respective image spaces) are given by
$C=(c_{ij})$ and the transposed matrix via
\begin{equation}\hat{C}:e_i'\to \sum_jc_{ij}e_j'=\phi_i\quad\text{etc.}\end{equation}

From linear algebra we know that the dimension of these subspaces is
given by the rank of the matrices $C,C^T$, with the ranks of $C$ and
its transpose $C^T$ being equal.  Identifying the matrix $C$ with the
corresponding abstract map $\hat{C}$, it is obvious that this
dimension cannot depend on the chosen basis. Or put differently, while
the subspaces themselves may change under a change of basis, the
dimension is a \tit{unitary invariant}. Without taking appropriate
measures, these subspaces may (in general) have an infinite or ``too
large'' dimension. What we are rather implying is the possibility to
find lower-dimensional \tit{essential subspaces}.
\begin{obdef}\label{dim}The dimension of the subspaces, spanned by $\phi_i$ or
  $\psi_j$ as image vectors under the maps $\hat{C}\,,\,\hat{C}^T$,
  applied to certain bases are equal and independent of the chosen
  bases in $\mcal{H}_1\,,\,\mcal{H}_2$. Furthermore the logarithm of
  the dimension is representing a measure of the entanglement
  entropy in leading order under certain favorable conditions. These
  conditions may be: All image vectors contribute with roughly the
  same strength, or, rather, their respective weights are not too
  different. Or, some of the matrix elements happen to be very small
  compared with the rest so that one can go over to a purified matrix;
  in particular, some of the image vectors may be very small and can
  be neglected. In all these cases one will get some approximation of
  the original entanglement-entropy.
\end{obdef} 
\begin{koro}Note that in the Schmidt-basis the dimension of the
  essential subspace can particularly easily be read of. It is the
  dimension of the subspace spanned by the eigenvectors of $W_1$ or
  $W_2$ with non-marginal eigenvalues $\lambda_i$. It can of course
  happen that there is no natural division between marginal and
  non-marginal eigenvalues. This is for example the case when the
  entanglement entropy is proportional to the volume, a situation we
  will observe in the following.
\end{koro}
In the following sections we will apply this approximative concept of
entanglement-entropy.

\section{Perturbation Theory of Hamiltonians and Entanglement}
\label{perturbation}
The preceding section shows clearly that the details of the dependence
of (entanglement-)entropy on volume and/or area as a function of the
type of global vectorstate are presumably subtle and intricate and
need more qualifications. This holds the more so if one wants to
discuss whole classes of models. Therefore, it is a natural idea to
study eigenstates of some Hamiltonian as described above. In most of
the examples we are aware of, groundstates of certain physical models
have been studied. It hence suggests itself to extend this
investigation and inspect both ground and excited states more closely
and try to infer characteristics of the respective entanglement
entropy from the (general) properties of such eigenstates.

A general method which suggests itself is perturbation theory of the
Hamiltonians under discussion. The strategy is the following. We
start from a Hilbert space, $\mcal{H}_V$, over a macroscopic volume
$V$ and divide it into two connected subvolumes, $V=V_1\cup V_2$, both
$V_1\,,\,V_2$ still being macroscopic with
\begin{equation}\mcal{H}_V= \mcal{H}_1\otimes \mcal{H}_2
\end{equation}
We assume a Hamiltonian, $H_V$, to be given on $\mcal{H}_V$ of the
kind described above. This Hamiltonian can be written as
\begin{equation}H_V=H_{V_1}+H_{V_2}+H_{bd}      \end{equation}
with $H_{V_i}=:H_i$ the commuting Hamiltonians of
the regions $V_1\,,\,V_2$ and $H_{bd}$ denoting the part of the
interaction which comprises lattice points of both $V_1$ and
$V_2$. That is, we have
\begin{equation}[H_1,H_2]=0\;,\;[H_i,H_{bd}]\neq 0      \end{equation}
It sometimes happens that we have to discriminate between, for
example, $H_1$ restricted to the subspace $\mcal{H}_1$ and its
embedded version, acting on the full Hilbert space by tensoring with
the unit operator of the volume $V_2$.
\begin{defi}We denote the embedded operators by $H_1$ etc. and the
  restricted versions by $H_1^r$ etc.
\end{defi}

The idea now is to proceed in the following way (we explain it for
the groundstate). We generally assume that our groundstates are not!
degenerate (this is frequently the case). We therefore have the unique
groundstate, $\Psi_0^{(0)}$, of our reference Hamiltonian, $H_1+H_2$,
which is the tensor product of the groundstates of $H_i$, i.e.
\begin{equation}\Psi_0^{(0)}=\psi_0^{(1)}\otimes \psi_0^{(2)}
\end{equation}
 One may now hope that one can get information about the
structure of the groundstate of the full Hamiltonian,
$H=H_1+H_2+H_{bd}$, by employing perturbation theory with respect to
$H_1+H_2$ around the reference groundstate $\Psi_0^{(0)}$. The problems
and obstacles which show up if one wants to follow this route are
described below. But anyway, as this strategy may not be entirely
futile and, to our knowledge, has not been attempted in the past in
this field, we describe now some of the necessary steps.

Generically $H_i$ are of ``size'' $V_i$ (e.g. their respective
operator norms or (most of their) eigenvalues). That is
\begin{equation}\parallel H_i\parallel\leq C_i\cdot n(V_i)=O(V_i)\;,\;\parallel
  H_{bd}\parallel\leq C'\cdot n(V_1,V_2)=O(boundary_{1,2})
\end{equation}
Therefore one may have the idea to treat $H_{bd}$ as a relatively
small perturbation of the operators $H_1$ or $H_2$.

The operators $H_1\,,\,H_2$, defined above over the regions
$V_1\,,\,V_2$, commute. In the following we will encounter in various
arguments such pieces of the total Hamiltonian which commute. Another
example is the following. We subdivide $V_1$ further into regions
$V_1'\,,\,V_1''$ and correspondingly for $V_2$ with $V_1=V_1'\cup
V_1'$. $V_1''$ is the region in $V_1$ which lies within distance
$d\geq\delta$ ($\delta$ the maximum over the ranges of the various
interaction potentials ) of the common boundary with $V_2$. In certain
calculations we choose $d$ macroscopic but $L\gg d\gg\delta$ (where,
as usual, we take $\delta$ as a microscopic quantity). The respective
Hilbert spaces are $\mcal{H}_1'\,,\,\mcal{H}_1''$ with
$\mcal{H}_1=\mcal{H}_1'\otimes\mcal{H}_1''$.  We can now define
another approximation of the total Hamiltonian $H$ in deleting the
boundary terms with respect to the interfaces separating
$V_1'\,,\,V_1''$ on the one hand and $V_2'\,,\,V_2''$ on the other
hand,
\begin{equation}\label{approx}H':=H_1'+H_1''+H_{bd}+ H_2'+H_2''=:H_1'+ H'_{bd}+ H_2'   \end{equation}
with
\begin{equation} H'_{bd}:= H_1''+H_{bd}+H_2''    \end{equation}
We now have
\begin{ob}The operators $H_1'\,,\,H'_{bd}\,,\,H_2'$ commute and
  \begin{equation}H=H'+H_{bd(1)}+H_{bd(2)} \end{equation} with the two
  boundary contributions describing the interaction through the
  interfaces between $V_1'\,,\,V_1''$ and $V_2'\,,\,V_2''$. The
  advantage is that we now still have included the interaction through
  the common interface between $V_1\,,\,V_2$ in $H'$, the interface we
  are originally interested in.
\end{ob}
As to such commuting operators we have the following spectral result
(cf. e.g. \cite{Achieser} or \cite{Riesz}) which goes back to v.Neumann.
\begin{satz}With (for simplicity reasons) $H_1,H_2,\ldots$ commuting
  bounded selfadjoint operators, they all are functions of a common
  selfadjoint operator, $A$, i.e. $H_i=f_i(A)$. It follows in
  particular that, in case the spectra are discrete, it exists a
  complete set of common eigenvectors for this set of commuting
  Hamiltonians (including multiplicities).
\end{satz}

One problem which however arises immediately if one wants to apply
perturbation theory of operators is the following (if one is not
entirely cavalier as to mathematical rigor). We know from almost every
discussion of the foundations of statistical mechanics that for
macroscopic volumes the spectrum of e.g. the corresponding
Hamiltonians, while being frequently discrete, is nevertheless so
extremely dense and/or highly degenerate that ordinary perturbation
theory is practically useless. A rough estimate yields the following
qualitative results. The number of eigenvalues (counting degeneracy)
of a hermitean matrix is the same as the dimension of the underlying
Hilbert space. This means in our case
\begin{equation}\#(eigenvalues\;of\;H_V)=D^{N_V} \end{equation} ($D$
the dimension of the local Hilbert spaces $\mcal{H}_{x_i}$, $N_V$ the
number of sites in $V$). On the other hand, the spectrum of the
corresponding Hamiltonian extends typically over an interval of order
$|V|$ .

That is, whereas the higher excited states are typically much more
degenerate and the spectrum is certainly not evenly distributed, a
very crude estimate yields a typical \tit{density of states} of the
order $O(|V|^{-1}\cdot D^{|V|})$. This prevents the immediate and
naive application of ordinary (analytic) perturbation theory, which
works well for perturbations which are small compared to the distance
of neigboring eigenvalues of the unperturbed Hamiltonian. To be more
precise, one knows from beautiful results derived by Rellich (see
\cite{Rellich}), and in particular for the finite dimensional case,
that for hermitean perturbations the (discrete) eigenvalues and
eigenstates are \tit{real-holomorphic} functions in the coupling
constant and that this does not only hold for very small values (see
also \cite{Baumgaertel}, \cite{Riesz} or \cite{Kato}).
\begin{satz}With $H_{\varepsilon}:= H_0+\varepsilon V$ selfadjoint for
  $\varepsilon\in \R$, $H_0\,,\,V$ bounded (for simplicity reasons)
  and $H_{\varepsilon}$ having purely discrete spectrum, the
  eigenvalues $\lambda_i(\varepsilon)$ and eigenvectors
  $\psi_i(\varepsilon)$ of $H_{\varepsilon}$, with
  $\lambda_i(0)\,,\,\psi_i(0)$ the eigenvalues and
  eigenvectors of $H_0$, are real analytic functions of
  $\varepsilon$. One can in particular choose $\varepsilon=1$. It can
  however happen that eigenvalues cross (and hence degeneracies
  change in a superficial sense; see the following corollary).
\end{satz}
\begin{koro}This has the important consequence that multiplicities
  belonging to a fixed $\lambda_i(\varepsilon)$ can only change at
  $\varepsilon=0$ as analytic functions, being identical on a certain
  interval, are necessarily the same everywhere. The other
  $\varepsilon$-values where a singular behavior can occur lie in the
  complex plane away from the real axis. Note however that, as the
  eigenvalue functions can cross at some points, the counting of
  degeneracies at such points is a matter of taste. The dimension of
  the total eigenspace is of course the dimension of the union of the
  individual eigenspaces belonging to the different
  $\lambda_i(\varepsilon)$ which meet at that point.
\end{koro}
Remark: The deeper reason why the nasty Puiseux-series can be avoided
derives from the fact that we have that $H_{\varepsilon}$ is
self-adjoint for real $\varepsilon$ (see \cite{Rellich}).\vspace{0.3cm}

On the other hand, convergence radii of the corresponding local power
series expansions happen to be of the order of the distances between
the points of the spectrum. This prevents to some extent concrete
quantitative estimates. To see more clearly the true nature of the
problem, we can for example start from the unperturbed groundstate,
$\Psi_0^{(0)}$, of $H^{(0)}:=H_1+H_2$ and try to infer with the help
of perturbation theory the structure of the corresponding groundstate
$\Psi_0$ of $H^{(0)}+H_{bd}$ as a power series expansion with respect
to the eigenvectors of $H^{(0)}$. That is,
\begin{equation}\Psi_0=\sum c_m\cdot \Psi_m^{(0)}    \end{equation}
with
\begin{equation}c_m=c_m^{(0)}+c_m^{(1)}+\ldots\;,\;c_0^{(0)}=1\,,\,c_m^{(0)}=0\;\text{for}\;m\neq 0\end{equation}
The first order yields
\begin{equation}c_m^{(1)}=V_{m0}/(E_m^{(0)}-E_0^{(0)})\;,\;m\neq
  0\,,\quad\text{and}\quad V_{m0}=(\Psi_m^{(0)}|V\cdot \Psi_0^{(0)})     \end{equation}
We see that for perturbation theory to make sense, 
\begin{equation}|V_{m0}|<|E_m^{(0)}-E_0^{(0)}|    \end{equation}
\begin{ob}While $H_{bd}$ is much smaller than $H_1$ or $H_2$ in
  general, it is still a macroscopic perturbation compared to the
  typically microscopic distances between the eigenvalues of $H_i$. So
  ordinary perturbation theory is not immediately applicable.
\end{ob}
Remark: There exists however a (complicated and tedeous) way to deal
with such problems to a certain extent (at least in the physics
literature); see \cite{Dawydow}.\vspace{0.3cm}

What will however better work is another important method of
estimating eigenvalues and their behavior under perturbations which
does not focus so much on the motion of individual eigenvalues under a
perturbation but rather makes more global and qualitative statements.
This method provides however no information about the respective
eigenvectors, our main point of interest. The method is based on the
so-called \tit{Rayleigh-Ritz-principle} and/or the
\tit{Poincare-Courant-Weyl estimates} (see \cite{Riesz},
\cite{Weinstein} or \cite{Bhatia}). All these statements are based on
\tit{minimum-maximum- or maximum-minimum-estimates} and the principle
of stronger or weaker constraints on sets of comparison Hilbert space
vectors.

A result, useful in our context, can e.g. be found in \cite{Riesz},
p.224, called the Weyl-Courant-inequalities, which we reformulate here
for bounded hermitean operators with discrete and only finitely
degenerated spectrum (not having zero as an accumulation
point).\\[0.3cm]
Remark: For various reasons the numbering of eigenvalues is different
in \cite{Riesz}. We start the counting, beginning with the
groundstate.
\begin{satz}With $A_1,A_2$ operators of the above kind, with sets of
  eigenvalues, chosen in increasing order (counting multiplicity),
\begin{equation}E_0^1\leq E_1^1\leq\ldots\;;\;E_0^2\leq E_1^2\leq\ldots\     \end{equation}
and 
\begin{equation}E_0\leq E_1\leq\ldots      \end{equation}
the corresponding eigenvalues of $A:=A_1+A_2$, we have the estimates
\begin{equation}E_{p+q}\geq E_p^1+E_q^2\;,\;p,q=0,1,2,\ldots     \end{equation}
\end{satz}
\begin{koro}With $H=H_0+V$ and $V$ a small perturbation of $H_0$, we
  have
\begin{equation}E_0^1-\parallel V\parallel\leq E_0\leq E_0^1+\parallel V\parallel      \end{equation}
and more generally
\begin{equation}E_p^1-\parallel V\parallel\leq E_p\leq E_p^1+\parallel V\parallel     \end{equation}
$E_0,E_0^1$ the groundstates of $H,H_0$ respectively.
\end{koro}
Proof: The lhs of the inequalities follow directly from the theorem
and $|E_q^2|\leq \parallel V\parallel$ for all the eigenvalues of
$V=A_2$. The rhs follows from the theorem by interchanging the roles
of the operators, that is
\begin{equation}A_1=A-A_2       \end{equation}
and hence
\begin{equation}E_{p+q}^1\geq E_p+E_q(-A_2)\geq E_p-\parallel A_2\parallel      \end{equation}
i.e.
\begin{equation}E_p\leq E_{p+q}^1+ \parallel A_2\parallel     \end{equation}
which yields the result by choosing $q=0$.\bewende
\section{The Groundstate of the Hamiltonian}
\label{groundstate}
We begin with the calculation of the entanglement-entropy of the
ground state, $\Psi_0$ of the full Hamiltonian over $V=V_1\cup V_2$.
In the following we use the leading-order identification of
entanglement-entropy made in observation/definition \ref{dim} and the
perturbational results of the preceding section.  In a first step we
study the entanglement-entropy of the ground state, $\Psi_0'$, of the
approximate Hamiltonian, $H'$, introduced in the preceding section (see
formula (\ref{approx})). We saw that $H'$ can be written as
\begin{equation}H'=H_1'+H_2'+H_{bd}'      \end{equation}
with all the terms on the rhs commuting with each other. Assuming
again that the ground states are not degenerate we infer from the
results of the previous section that the ground state energy, $E_0'$,
of $H'$ can be uniquely written as
\begin{equation}E_0'=E'_{0,1}+E'_{0,2}+E'_{0,bd}     \end{equation}
with the rhs the sum of the ground state energies of the terms
occurring on the rhs of the previous equation.\\[0.3cm]
Remark: Note that the embedded Hamiltonians always have full
subspaces, belonging to an eigenvalue. For example, $H'_1$ has the
eigenspace $\psi'_{0,1}\otimes \mcal{H}''_1$ belonging to the ground
state energy $E'_{0,1}$ in the Hilbert space
$\mcal{H}_1=\mcal{H}'_1\otimes\mcal{H}''_1$ and $\psi'_{0,1}$ the
unique ground state of the restricted $H'_{1,(r)}$ acting on
$\mcal{H}'_1$.\\[0.3cm]
The need to constantly make these distinctions is a bit nasty and we 
will be a little bit sloppy if no confusion can arise. We then have
\begin{ob}The ground state, $\Psi'_0$ of $H'$ can now be uniquely
  represented as the tensor product of the ground states of the
  restricted Hamiltonians, i.e.
\begin{equation}\Psi'_0=\psi'_{0,1}\otimes \psi'_{0,2}\otimes \psi'_{0,bd}     \end{equation}
where
\begin{equation}\psi'_{0,bd}\in \mcal{H}''_1\otimes  \mcal{H}''_2   \end{equation}
\end{ob}

In order to calculate the partial traces with respect to $\mcal{H}_1$
or $\mcal{H}_2$ we have in a first step to develop $\Psi'_0$ with
respect to a basis of $\mcal{H}_1\otimes\mcal{H}_2$ or, what amounts
to the same,
$\mcal{H}'_1\otimes\mcal{H}''_1\otimes\mcal{H}'_2\otimes\mcal{H}''_2$.
Choosing as bases in the subspaces the eigenvectors of the restricted
Hamiltonians
$H_{1,(r)}'\,,\,H_{2,(r)}'\,,\,H_{1,(r)}''\,,\,H_{2,(r)}''$, we can
infer the following from the above observation.
\begin{conclusion}In the representation of $\Psi'_0$ with respect to
  the mentioned basis in 
\begin{equation}\mcal{H}_1\otimes\mcal{H}_2=
  \mcal{H}'_1\otimes\mcal{H}''_1\otimes\mcal{H}''_2\otimes\mcal{H}'_2
\end{equation}
the outermost left and right terms on the rhs remain fixed (no
summation). The only summation occurs in the boundary term,
$\psi'_{0,bd}$, which is developed with respect to a basis in
$\mcal{H}''_1\otimes\mcal{H}''_2$. Taking for example the eigenvectors
of $H_{1,(r)}''\,,\,H_{2,(r)}''$ we write
\begin{equation}\psi_{0,bd}'=\sum c'_{i,j}\,\psi''_{i,1}\otimes\psi''_{j,2}   \end{equation}
and
\begin{equation}\Psi_0'=\psi_{0,1}'\otimes(\sum\ldots)\otimes\psi'_{0,2}    \end{equation}
\end{conclusion}
\vspace{0.3cm}

We have 
\begin{equation}H'_{bd}=H_1''+H''_2+H_{bd} \end{equation} where the
operators occurring on the rhs are all of roughly the same size, i.e.
of order $O(boundary_{1,2})$. While $H_1''\,,\,H''_2$ commute, the
support of $H_{bd}$ overlaps both with the support of $H_1''$ and
$H''_2$ and the respective commutators are typically different
from zero.

If we now view $\Psi'_0$ as a state over $\mcal{A}^1$, the algebra on
$\mcal{H}_1=\mcal{H}'_1\otimes \mcal{H}''_1$ we get:
\begin{equation}(\Psi'_0|(A\otimes\mbf{1})\Psi'_0)=(\psi_{0,1}'\otimes
  \psi_{0,bd}'|(A\otimes\mbf{1})\psi_{0,1}'\otimes \psi_{0,bd}')
\end{equation}
 where on the lhs $\mbf{1}$ is the unit operator on
$\mcal{H}_2$, on the rhs it denotes the unit operator on
$\mcal{H}''_2$. Inserting $\psi_{0,bd}'=\sum
c'_{i,j}\,\psi''_{i,1}\otimes\psi''_{j,2}$ in the above expression we
get
\begin{multline}(\Psi'_0|(A\otimes\mbf{1})\Psi'_0)=\sum_{ij}\sum_l
  \overline{c}'_{il}c'_{jl}(\psi'_{0,1}\otimes \psi''_{i,1}|A\circ
  \psi'_{0,1}\otimes \psi''_{j,1})\\
=\sum_{ij}b_{ij}(\psi'_{0,1}\otimes \psi''_{i,1}|A\circ\psi'_{0,1}\otimes \psi''_{j,1})
    \end{multline}
with $b_{ij}=\sum_l\overline{c}'_{il}c'_{jl}$.
\begin{conclusion}The reduced state or density matrix on $\mcal{H}_1$,
  corresponding to the total vector state, $\Psi_0'$, is
\begin{equation}W_1'=|\psi'_{0,1}>< \psi'_{0,1}|\otimes W_1''   \end{equation}
with $W''_1$ the density matrix on $\mcal{H}''_1$ with matrix elements
$b_{ij}$.
\end{conclusion}

If the local Hilbert spaces have uniform dimension $D$ and with the
assumed finite interaction distance $\delta$, we conclude that the
dimension of the Hilbert space $\mcal{H}''_1$ is of order
$O(D^{|bd_{1,2}|})$. From the preceding conclusion we infer that for
the vector $\Psi'_0$ all perturbations are essentially restricted to
the boundary region. 
\begin{conclusion}For lattice Hamiltonians as we have introduced them,
  the groundstate of $H'$ is expected to contain or is scattered over
  a number of eigenstates of $H''_1$ of the order $O(D^{|bd_{1,2}|})$.
  Correspondingly we infer that its entanglement-entropy is of order
  $O(|bd_{1,2}|)$.
\end{conclusion}
Remark: The deeper reason why we are able to infer such a general
result for the approximate Hamiltonian $H'$ is, on the one hand, the
sufficient localisation of the perturbation in a boundary layer of
finite thickness and, on the other hand, the uniqueness properties of
the groundstate as a tensor product of the corresponding groundstates
of the Hamiltonians of the subvolumes.\vspace{0.3cm}

Now we come to the groundstate of the full Hamiltonian 
\begin{equation}H=H'+H_{bd(1)}+H_{bd(2)} \end{equation} 
The difference
between $H$ and $H'$ is a small perturbation on the scale of $H$ or
$H'$ as operators, but not on the scale defined by the
difference between neighboring eigenvalues of $H\,,\,H'$. From the
preceding section we know at least that
\begin{equation}E'_P-\triangle\leq E_p\leq E'_p+\triangle \end{equation}
with 
\begin{equation}\triangle=\parallel H_{bd(1)}+H_{bd(2)}\parallel=O(|bondary_{1,2}|) \end{equation}
Our plan is to make an inference from the number of eigenstates of $H_1'+H_1''$ which
essentially contribute in the representation of $\Psi_0'$ to the
corresponding number which essentially contribute in the
representation of $\Psi_0$, the groundstate of the full
Hamiltonian. This number should be of the same order as the
corresponding number of eigenstates of $H_1$, as both sets represent
complete bases in $\mcal{H}_1=\mcal{H}'_1\otimes \mcal{H}''_1$.

We do this in several steps employing the following reasoning. In a first step
we add the boundary interaction $H_{bd(2)}$ to the start Hamiltonian
$H'$ yielding the intermediate Hamiltonian $H'_1+H(V''_1\cup
V_2)$. Its groundstate is $\psi'_{0,1}\otimes \phi'_0$ with $\phi'_0$
the groundstate of $H(V''_1\cup V_2)$. In $V_1$ we have more or
less the same situation as before with possible perturbations again
confined (by definition) to the region $V_1''$. The same argument as
before yields an entropy for the reduced state over $V_1$ of order
$O(|bd_{1,2}|)$. Now we employ the fact that the entropies are
necessarily the same on both sides. That is, we arrive at
\begin{ob}The entropy of the state $\phi'_0$ reduced to $V_2$ is of
  order $O(|bd_{1,2}|)$.
\end{ob}

Now we employ the localisation properties of $H_{bd(1)}$ about the
interface $bd_1$ within a small strip of diameter $2\delta$, the
interface itself having distance $d\gg\delta$ from the common boundary
between $V_1$ and $V_2$. From general experience, drawn from the
foundations of statistical mechanics and many-body-theory, we feel
allowed to assume that deep inside the region $V_2$, i.e. in $V'_2$,
the groundstate, $\Psi_0$, of the full Hamiltonian $H$ should look
similar to the groundstate, $\Phi'_0$ of the Hamiltonian $H(V''_1\cup
V_2)$.\\[0.3cm]
Remark: Note that we always make the assumption (cf. the introduction)
that our system is not in a \tit{quantum-critical state}, i.e.
correlations do not extend to infinity. The latter case would need
some extra discussion.  \vspace{0.3cm}

Concerning the groundstate of the latter Hamiltonian we learned that
its restriction to $V_2$ has an entropy  of order $O(bd_{12})$. From
the above we again conclude that, in $V_2$, $\Psi_0$ differs from
$\Psi'_0$ (the groundstate of $H'$) essentially in a boundary layer
about the interface $bd_{12}$. By symmetry we infer the same for the
region $V_1$ and arrive at
\begin{conclusion}The above chain of reasoning leads to the conclusion
  that as a consequence of the spatial localisation properties of
  $H_{bd}\,,\,H_{bd(1)}\,,\,H_{bd(2)}$ and certain natural assumptions
  about clustering or decay of influence and/or interactions, the
  groundstate of the full Hamiltonian, $H$, has an
  entanglement-entropy of order $O(bd_{12})$.
\end{conclusion} 
Remark:This general conclusion is corroborated by exact results for
(typically) low-dimensional models; cf. e.g. the literature mentioned
in the introduction.  

\section{\label{highly}The Higly Excited Eigenstates}
It turns out that for highly excited eigenstates it is more fruitful
to adopt a completely different strategy, which is strongly inspired
by ideas and methods taken from statistical mechanics. To simplify
the discussion we treat the following system. We take a huge box of
sidelength $L$ as the total volume $V$.  We partition it by a lattice
of small boxes, $C_i$, of sidelength $l$ with $L\gg l\gg \delta$
($\delta$ the range of the interaction in the original Hamiltonian,
$H$. I.e., we assume that $l$ is small but still macroscopic; this is
the usual assumption in statistical mechanics. As subvolumes,
$V_1\,,\,V_2$ we take certain regions in $V$ each of which contains an
integer number of such small boxes. I.e., we assume (with
$N_l\,,\,N_{l,1}\,,\,N_{l,2}$ the respective numbers of boxes in
$V\,,\,V_i$)
\begin{equation}N_l=N_{l,1}+N_{l,2}\;,\;N_l=L^3/l^3\,,\,|V_i|=l^3\cdot
  N_{l,i} \end{equation}
 Each of the small boxes contains $l^3/a^3$
lattice sites of the original lattice. We assume of course that the
interface, separating $V_1$ and $V_2$, is sufficiently regular, i.e.
its area is assumed to be of order $O(L^2)$.

In each of the small boxes, $C_i$, we take as Hamiltonian, $h_i$, the
piece of our original total Hamiltonian $H$ with interaction terms
confined to $C_i$, i.e., leaving out the interaction terms occurring
in $H$ between the different boxes. Note that, due to the assumed
translation invariance of our interaction, all the $h_i$ are
equivalent as operators. As these small boxes still contain quite a
few lattice sites, the spectrum of $h=h_i$ may still beboth complex
and degenerated. As new reference Hamiltonians in $V\,,\,V_j$, $j=1,2$, we
take
\begin{equation}H':=H'_V:=\sum_V h_i\;,\;H'_j:=\sum_{V_j}h_i \end{equation}

There is now, in contrast to the preceding section, no boundary term
$H_{bd}$, operating in the vicinity of the boundary, $bd_{12}$, but
the entanglement structure may still be quite complex as we will see
below.\\[0.3cm]
Remark: Such truncated systems are frequently discussed and their
properties exploited in quantum statistical mechanics within the
Gibbsian (ensemble) approach. See for example
\cite{Becker},\cite{Tolman},\cite{Schroedinger} or
\cite{Fowler}.\\[0.3cm]
As all these $h_i$ commute (by construction), the eigenstates and
eigenvalues of $H'$ or $H'_j$ can be built up from the more
elementary components belonging to the $h_i$.

So, to begin with, let us start from some macroscopic volume, $V'$, of
the above kind ($V'=V$ or $V_j$), the Hamiltonian,
$H'=\sum_{V'}h_i$, and some eigenvalue, $E$, sufficiently far away
from the groundstate energy $E_0=\sum E_0^{(i)}$ with $E_j^{(i)}$
denoting the j-th energy level of the box Hamiltonian $h_i$. We have
\begin{equation}E=\sum_{C_i}E_j^{(i)} \end{equation} for certain
appropriate combinations of energy levels, $E_j^{(i)}$, of the
$h_i$. It is here where certain arguments of combinatorial statistical
mechanics enter.

The problem can now be phrased a little bit differently. With $N$
boxes given and $h=h_i$ having the energy levels (counting multiplicity!)
$E_1\leq E_2\leq\ldots\leq E_j,\ldots$, we are interested in the number of ways
of distributing the energies, $E_j$, over the $N$ boxes under the
constraints
\begin{equation}N=\sum N_j\quad,\quad E=\sum N_j\cdot E_j  \end{equation}
with $N_j$ the number of boxes having energy $E_j$. Each such
configuration is hence characterized by the sequence,
$(N_1,N_2,\ldots,N_j,\ldots)$. We then have
\begin{ob}To each fixed configuration $(N_1,N_2,\ldots,N_j,\ldots)$
  the number of ways of distributing the energies $E_j$ over the $N$
  boxes under the above constraints is
\begin{equation}W=(N!/N_1!\cdot\cdots N_j!\cdots)\quad,\quad N=\sum N_j\;,\;E=\sum N_jE_j  \end{equation}
\end{ob}

From the combinatorics of such expressions one knows that there exists
a pronounced maximum of $W$ for a special configuration
$(N_1,N_2,\ldots,N_j,\ldots)_{max}$ (cf. the above cited literature
for more details). The constraints can be implemented via Lagrange
multipliers with, in the end, the multiplier $\beta$, belonging to the
$E$-constraint, turning out to be something like an \tit{inverse
  temperature}. It is however important (while usually not openly
mentioned in the literature) that for this and other results to hold,
$(E-E_0)$ has to be \tit{macroscopic}, i.e. of order $O(|V|)$. This
implies that, with the individual levels, $E_j$ of $h$, being
microscopic, most of the occurring $N_j$ are sufficiently large so
that \tit{Stirlings formula} can be applied. After some calculations
one winds up with the formula (for details see e.g. \cite{Tolman},
chapt. 87, p.364 ff, or \cite{Becker}, chapt. 46, p. 150 ff)
\begin{conclusion}\label{config}Under the assumptions being made we
  have for the most probable configuration and the configuration
  entropy:
\begin{equation}
  N_j/N=e^{-\beta E_j}/\sum e^{-\beta E_j}\;,\;
  \ln W_{max}=N\cdot\ln (\sum_j e^{-\beta
    E_j}+\beta\cdot E)  \end{equation}
(with $\beta$ only implicitly given by the first equation). In any case, $\ln W_{max}$ turns
out to be in general proportional to the volume $|V|$ ($N\sim |V| $)
for highly excited states.
\end{conclusion}

These findings have the following consequences for our entanglement
problem. With
\begin{equation}\Psi_E=\sum c_{ij}\,\phi_i^{(E_1)}\otimes
  \psi_j^{(E_2)}\quad,\quad E_1+E_2=E  \end{equation}
and $\phi_i^{(E_1)}\,,\,\psi_j^{(E_2)}$ eigenvectors to the fixed
energies $E_1\,,\,E_2$ of $H'_1\,,\,H'_2$, respectively, this is an
eigenvector for $H'$ with energy $E$. We have just seen that the
eigenvalues $E_1$ and $E_2$ are extremely degenerated so that the
number of terms, occurring in the above sum, can in principle be
chosen  to be of order $O(e^V)$. 

 Note that, in addition, we could
also sum over all possible combinations of $E_1\,,\,E_2$ with
$E_1+E_2=E$ but this is not necessary for our argument. We can now
make various choices. We can for example select a very special and simple
eigenvector of product type (i.e. all $c_{ij}=0$ except one):
\begin{equation}\Psi_E= \phi^{(E_1)}\otimes\psi^{(E_2)}\quad,\quad
  E_1+E_2=E \end{equation}
Its entanglement-entropy is of course zero.

On the other hand, due to the huge degeneracy of all macroscopic
energy levels of $H'$, we can exploit our above conclusion
\ref{config} and what we said in the preceding sections about our
coarse approximation of entanglement-entropy. It is easy to choose the
$c_{ij}$ almost evenly spread over the full range of degenerated
eigenstates belonging to $E_1$ and $E_2$ in such a way that a typical
eigenvector to energy $E$ has an entanglement-entropy which is
proportional to the volume $|V|$. That is
\begin{conclusion}Due to the huge degeneracy of macroscopically
  excited energy levels of $H'$, the typical eigenvector, belonging to
  the class of eigenvectors of such an energy level, has an
  entanglement-entropy of order $O(|V|)$, more specifically
  \begin{equation}S(W_1)=\alpha_1\cdot |V_1|= \alpha_2\cdot
    |V_2|=S(W_2)\end{equation} (a consequence of conclusion
  \ref{config}).
 Furthermore, our preceding discussion shows that
  these states, $W_1\,,\,W_2$ display features we know from
  statistical mechanics. By ``typical'' we mean, by randomly
  selecting one of the admissible eigenvectors from the huge class, we
  will get such a state
  with high probability as one knows that the combinatorial
  coefficients, calculated in conclusion \ref{config}, are extremely large for
  macroscopic volumes $V,V_j$.
\end{conclusion}
\begin{ob} We remind the reader of our construction of a vector state
  belonging to a canonical equilibrium state of system $(1)$ with the
  help of tensoring with a system $(2)$. Our above findings on higly
  excited states represent, so to speak, the dual version of this
  observation. Highly excited states on $V$ have, as we have seen, a
  tendency to resemble states on, say, $V_1$ which display a marked
  statistical mechanical behavior (they are of course not always true
  equilibrium states).
\end{ob}

One can now go on and study the full untruncated Hamiltonian, $H$,
starting from such a reference Hamiltonian $H'$. Assuming that by
inserting the usually very small boundary terms between the blocks
$C_i$ and adding the boundary Hamiltonian, $H_{bd}$, nothing
spectacular will happen (the ordinary assumption in statistical
mechanics), the degeneracy of eigenvalues and/or the density of states
will remain essentially the same. This may be inferred from
perturbation theory. We therefore arrive at the final conclusion
\begin{conclusion}In the case of the full Hamiltonian $H$, we have
  essentially the same result as in the preceding conclusion for the
  reference Hamiltonian $H'$. Due to the expected huge density of
  states and/or the huge degeneracy of eigenvalues we have for
  macroscopically excited eigenvalues, $E$, that a randomly selected (i.e.,
  generic) eigenstate has an entanglement-entropy which is
  proportional to the volume in leading order.
\end{conclusion} 
\section{Low-Lying Excited States}
We now discuss the special case that the excited states lie in the
vicinity of the groundstate, i.e. instead of energy levels fulfilling
\begin{equation}\triangle :=(E-E_0)= O(|V|)    \end{equation}
we deal with excitation energies which are much smaller. It is perhaps
surprising that in this case we have to use yet another strategy. The same
general formula
\begin{equation}W=(N!/N_1!\cdots N_j!\cdots) \end{equation} holds of
course also in this regime but for example Stirlings approximation is
no longer applicable as the $N_j$ are in general too small. Even if we
would ignore this fact (which would presumably only affect the
quantitative aspects of some estimates) there exists yet another more
serious problem. The energy constraint (we now denote the energy
levels of $H'$ by $E_i'$)
\begin{equation}\sum E_j\cdot N_j=E_0'+\triangle =N\cdot E_0+\triangle    \end{equation}
with $\triangle\ll |V|$, is more difficult to implement in this
regime. It is interesting to analyse the consequences of
$\triangle\ll |V|$. 

At first glance it seems that we will get the same results as in the
previous section by applying the same methods (and in the statistical
mechanics literature known to us we have found almost no remark as to
possible problems). The method we applied previously is
indeed very general but there exists a subtle point. The Lagrange
multiplier $\beta$ is only implicitly defined via the constraint
\begin{equation}E'/N=\sum E_j\cdot e^{-\beta E_j}/\sum e^{-\beta E_j}
\end{equation} i.e., it regulates the average energy per box
Hamiltonian, $h_i$, in form of a canonical distribution over the
energy levels of $h$. The $E_j$ are in general not known in detail but
one may infer that with $(E'-E'_0)=O(|V|)$ both sides are of the same
order for \tit{finite} $\beta$ so that it is reasonable that we can
find some definite value for which the implicit equation for $\beta$ can be
fulfilled. But we now have $(E'-E_0')/N\ll 1$ and we conjecture that in
this regime the above implicit equation can only be fulfilled for
$\beta\gg 1$ or $\beta\to \infty$ (which seems to be quite natural,
given the obvious similarities to statistical mechanics. A
``thermal'' state near the groundstate has by definition a low temperature).
\begin{ob}For $(E'-E_0')$ small, i.e. $(E'-E_0')\ll O(|V|)$, the
  parameter $\beta$ becomes very large. For these values it becomes
  difficult to reliably estimate the terms in the occurring variational
  equations which are now combinations of very large and very small
  terms. Note in particular that for $\beta$ large
\begin{equation}N_j=N\cdot e^{-\beta E_j}/\sum e^{-\beta E_j}      \end{equation}
 becomes very small compared to $N$.
\end{ob}

Therefore we choose another strategy which is better adapted to this
situation. We simply catalogue the low-lying excitations of $H'$
directly, beginning with the groundstate. We have
\begin{ob}1) For the groundstate we have: 
\begin{equation}E_0'=N\cdot E_0\;,\;\text{no degeneration}     \end{equation}
2) For one box hamiltonian excited we have:
\begin{equation}E_i'=(N-1)E_0+E_i\;,\;W(E_i')=N     \end{equation}
3) Two levels excited; there are two possibilities, $E_i=E_j$ or
$E_i\neq E_j$. We have
\begin{equation}E'_{ii}=(N-2)E_0+2E_i\;,\,W(E'_{ii})=(N\cdot
  (N-1)/2)     \end{equation}
or
\begin{equation}E'_{ij}=E_i+E_j+(N-2)E_0\;,\;W(E'_{ij})=N(N-1)   \end{equation}
etc.
\end{ob}
Remark: Note that these results of course coincide with the general
formula, if we insert the corresponding $N_i$.\vspace{0.3cm}

We see the following. Already for the lowest excited levels of $H'$ we
have a degeneracy, $W(E_i')=O(|V|)$ or a small power of $|V|$.
Repeating our previous arguments we infer
\begin{conclusion}Already the lowest excited levels of $H'$ have a
  degeneracy of order $O(|V|)$ or a small power of $|V|$, entailing
  that we can construct corresponding eigenstates having an
  entanglement-entropy of order $O(\ln |V|)$. Note that $H'$ does not
  contain the boundary term $H_{bd}$. For the full $H$ we hence expect
  a generic entanglement-entropy of order $O(\ln|V|\cdot |bd_{12}|)$
  where the additional factor $|bd_{12}|$ comes from the term $H_{bd}$
  as this term leads to a further splitting proportional to the area
  of the boundary as was for example the case for the groundstate. 
\end{conclusion}

The situation changes slightly if we go over to higher excited levels.
For, say, $k$ levels excited the two extreme cases are: 1) all $k$
levels identical or, 2) all levels being different. The intermediate
class comprises cases where some of the $E_j$ coincide. We have the
following estimate
\begin{koro}If $k$ levels are excited, $E_{i_1},\ldots
  ,E_{i_k}$, with repetitions allowed, we have the following estimate
  with respect to the degeneracies of the low-lying eigenvalues of $H'$
\begin{equation}(N!/k!\cdot (N-k)!)\leq W_k\leq
  (N\cdot (N-1)\cdots (N-k+1))     \end{equation}
For $N$ still very large compared to $k$, this entails
\begin{equation}W_k=O(N^k)\quad\text{and}\quad \ln W_k=O(k\cdot\ln N)= O(k\cdot\ln |V|)     \end{equation}
For the full Hamiltonian $H$ we again get an additional multiplicative
factor of order $O(|bd_{12}|)$.
\end{koro} 

\section{Appendix}
\begin{appendix}

\section{\label{infinite} Infinite-Dimensional Local Hilbert-spaces}
We assume that the spectra of the local and global hamiltonians are
discrete but are now \tit{not} necessarily bounded from above. We
still assume that $H_V$ is bounded from below (existence of a
groundstate!).  So let $H$ be such a Hamiltonian on a separable
Hilbert space, $\mcal{H}$. We select a certain (countable) basis,
$e_i$, and choose certain subspaces, $\mcal{H}_n$, spanned by the
basis vectors, $e_1,\ldots ,e_n$. The projector on $\mcal{H}_n$ is
denoted by $P_n$. Then
\begin{equation} P_n\cdot H\cdot P_n      \end{equation}
is a bounded operator on  $\mcal{H}_n$. Being a little bit more
general, if $H_V$ is the Hamiltonian on
$\mcal{H}_V:=\mcal{H}_{i_1}\otimes\cdots\otimes \mcal{H}_{i_N}$, we
select in each $\mcal{H}_{i_{\nu}}$ the subspace $\mcal{H}_i^{(n)}$,
spanned by $e^{(i)}_1,\ldots ,e^{(i)}_n$. From these local pieces we compose
the subspace, $\mcal{H}_n$, in $\mcal{H}_V$, i.e.
\begin{equation}\mcal{H}_n:= \mcal{H}_{i_1}^{(n)}\otimes\cdots\otimes \mcal{H}_{iN}^{(n)}   \end{equation}
and denote again the projector on this subspace by $P_n$. In the same
manner we take
\begin{equation}H^{(n)}:=P_n\cdot H_V\cdot P_n     \end{equation}
as finite dimensional Hamiltonian on these
$\mcal{H}_n\subset\mcal{H}_V $.

Now we have to discuss what happens if we take the limits
\begin{equation}\mcal{H}_n\to  \mcal{H}_V\;,\;H^{(n)}\to H_V\quad\text{etc.}   \end{equation}
Note that by construction all the $H^{(n)}$ are now bounded, even finite
dimensional, but the limit Hamiltonians are in general unbounded. 
We do not want to be too tedious concerning technical details of
functional analysis at this point. Suffice it to say that a reasonable
concept of operator convergence in this context is convergence in the
\tit{resolvent sense}, i.e., instead of dealing with unbounded
operators we deal with their bounded resolvents
\begin{equation}(H-z)^{-1}\;,\;\text{Im}(z)\neq 0     \end{equation}

Under quite weak assumptions, which are in general fulfilled in our
context, (strong) resolvent convergence can be assumed
(cf. \cite{Reed}, sect. VIII.7). As a consequence we have the
following result (discrete spectrum):
\begin{ob}With $H$ having discrete spectrum and if $H_n\to H$ in
  strong resolvent sense, we can find to each eigenvalue $E$ of $H$ an
  interval $(a,b)$ so that $E$ is the only spectral value of $H$ in
  $(a,b)$. With $P^{(n)}_{(a,b)}$ the spectral projections of $H_n$ on
  the interval $(a,b)$ we have
\begin{equation}P^{(n)}_{(a,b)}\psi\to P_{(a,b)}\psi=P_E\psi    \end{equation}
for all $\psi\in \mcal{H}$. We get even stronger results if we assume
convergence in norm-resolvent sense.
\end{ob}

We now sketch how results about entanglement-entropy, derived for
large but nevertheless finite dimensional systems, could be
transferred to the general case. But as there are several quite
delicate technical steps involved, which to rigorously prove would
need quite an amount of mathematical input, we refrain from giving all
the intricate mathematical details at the moment. From a physical
point of view the strategy seems to be quite reasonable. 

In a first step we have to guarantee that for example in the case of
groundstates
\begin{equation}\Psi_0^{(n)}\to \Psi_0      \end{equation}
In the following sections we regard these vectorstates as states on a
restricted region, $V_1$, and the corresponding Hilbert space or
observable algebra. We then have for the respective density matrices
over $\mcal{H}_1$ 
\begin{equation}W_1^{(n)}\to W_1     \end{equation}
in the form
\begin{equation}Tr(W_1^{(n)}\cdot A)\to Tr(W_1\cdot A)     \end{equation}
What we need is a result like
\begin{equation} W_1^{(n)}\cdot \ln W_1^{(n)}\to W_1\cdot \ln W_1     \end{equation}
in a suitable topology so that we may get in the end
\begin{equation}Tr( W_1^{(n)}\cdot \ln W_1^{(n)})\to
Tr(W_1\cdot \ln W_1)    \end{equation}
that is
\begin{equation}S(W_1^{(n)})\to S(W_1)    \end{equation}
Remark: Some of the necessary technical arsenal can be found for
example in \cite{Thirring}, see also the seminal paper by Wehrl
(\cite{Wehrl}), in particular section IID about \tit{continuity
  properties} of entropy as a function defined over the density
matrices, or the more recent \cite{Plenio2}. \vspace{0.3cm}

Note that in general the entropy is not continuous if the
density matrices are equipped with the trace-norm
\begin{equation}\parallel\rho_1-\rho_2\parallel_{tr}:=tr(|\rho_1-\rho_2|)
\end{equation}
It is an important observation that this property can be restored if
one works with the subset of density matrices having \tit{finite
  energy}.  This means, one chooses a fixed reasonable Hamiltonian,
$H$, so that the canonical Gibbs state, $\sigma_{\beta}$, at some
inverse temperature, $\beta$, has finite trace and finite \tit{mean
  energy}
\begin{equation}tr(\sigma_{\beta}\cdot H)=E            \end{equation}
The subset of admissible density matrices is then given by the
condition
\begin{equation}tr(\rho\cdot H)\leq E        \end{equation}
This is a reasonable condition for e.g. entanglement entropy. Taking
for the bipartite system a Hamiltonian of the form $H=H_1+H_2$
(i.e. no interaction across the common boundary) with reasonable
$H_i$, it suffices that the $H_i$ are in the domain of definition of
the pure state $\Psi$ as we then have
\begin{equation}tr(W_i\cdot H_i)=(\Psi,(H_i\otimes\mbf{1})\Psi)=E'       \end{equation}
for some $E'$. So, adapting $\beta$, these particular density matrices
always lie in appropriate subsets. 

\section{\label{entanglement}Proof of Theorem \ref{trace}}

Viewing the matrices $C:=(c_{jm})\,,\,C^*:=(\overline{C}^T$ as
operators from $\mcal{H}_2\rightarrow
\mcal{H}_1\,,\,\mcal{H}_1\rightarrow \mcal{H}_2$ respectively, we have
\begin{equation}W_1=C\cdot C^*\;,\;W_2=C^*\cdot C   \end{equation}
If $\psi_i$ is an eigenvector of $W_2=C^*\cdot C$, i.e.
\begin{equation}C^*\cdot C\circ\psi_i=\lambda_i\cdot\psi_i   \end{equation}
it follows that 
\begin{equation}CC^*C\circ\psi_i=\lambda_i\cdot C\circ\psi  \end{equation}
i.e., $\lambda_i$ is eigenvalue of $W_1=CC^*$ with eigenvector
$C\psi_i$ and vice versa. Furthermore, if $\lambda_i\neq 0$, the
degeneracy is the same with respect to $C^*C$ and $CC^*$. This follows
from 
\begin{equation}(C\psi^1_i|C\psi^2_i)=(\psi^1_i|C^*C\psi^2_i)=\lambda_i(\psi^1_i|\psi^2_i)   \end{equation}
that is, with $\psi^1_i\,,\,\psi^2_i$ orthogonal eigenvectors to the
eigenvalue $\lambda_i\neq 0$, $C\psi^1_i\,,\,C\psi^2_i$ are also
orthogonal and non-vanishing and the same holds in the other
direction.

By the same token we infer from the positive definiteness of $C^*C$ and
$CC^*$ that all eigenvalues are $\geq 0$. The normalisation of the
vector $\Phi$ as a state on $\mcal{A}$ and $\mcal{A}_i$ implies that
the trace norm of $W_i$ is one. This proves the theorem.\\[0.5cm]
{\small Acknowledgement: I thank the referees for various helpful comments.}

\end{appendix}

\begin
{thebibliography}{99}
  {\small
\bibitem{Jacobson}T.Jacobson,D.Marolf,C.Rovelli: ``Black Hole Entropy:
  Inside or Out?'', Int.J.Theor.Phys. 44(2005)1807, hep-th/0501103
\bibitem{Wald}R.M.Wald: ``The Thermodynamics of Black Holes'', Living
  Rev. Relativity, 4(2001), 6
\bibitem{Sorkin}R.D.Sorkin: ``Ten Theses on Black Hole Entropy'',
  Stud.Hist.Philos.Mod.Phys. 36(2005)291, hep-th/0504037

\bibitem{Bousso}R.Bousso: ``The Holographic Principle'',
  Rev.Mod.Phys. 74(2002)825, hep-th/0203101
\bibitem{Requ}M.Requardt: ``Planck Fluctuations, Measurement
  Uncertainties, and the Holographic Principle, Mod.Phys.Lett. A 22(2007)791, gr-qc/0505019, 
\bibitem{BH}M.Requardt: ``The Statistical Mechanics of Long-Range
  Bulk-Boundary Dependence on the Microscopic Scale in BH-Physics and
  Holography'', Preprint, Göttingen July 2007

\bibitem{Brustein1}A.Yarom,R.Brustein: ``Area-Scaling of quantum
fluctuations'', Nucl.Phys.B 709(2005)391, hep-th/0401081
\bibitem{Brustein2}A.Yarom,R.Brustein: ``Dimensional Reduction from
  Entanglement in Minkowski Space'', JHEP 0501(2005)46, hep-th/0302186
\bibitem{Requ1}M.Requardt: ``Symmetry Conservation and Integrals over
  Local Charge Densities'', Comm.Math.Phys. 50(1976)259
\bibitem{Requ2}M.Requardt: ``Fluctuation Operators and Spontaneous
  Symmetry Breaking'', J.Math.Phys. 43(2002)351, math-ph/0003012
\bibitem{Bombelli}L.Bombelli,K.Kaul,J.Lee,R.Sorkin: ``Quantum Source
of Entropy for Black Holes'', Phys.Rev. D34(1986)373
\bibitem{Srednicki}M.Srednicki: ``Entropy and Area'',
  Phys.Rev.Lett. 71(1993)666, hep-th/9303048
\bibitem{Plenio1}M.B.Plenio,J.Eisert,J.Dreißig,M.Cramer:
  ``Entropy, entanglement, and area'', PRL 94(2005)060503, quant-ph/0405142

\bibitem{Schroer}B.Schroer: ``More on Area Density of
  Localisation-Entropy and Problematization of Black Hole Entropy'', hep-th/0511291
\bibitem{Fazio}L.Amico,R.Fazio,A.Osterloh,V.Vedral: ``Entanglement in
  Many-Body Systems'', quant-ph/0703044
\bibitem{Zanardi}P.Zanardi: ``Quantum entanglement in fermionic
  lattices'', Phys.Rev. A 65(2002)042101
\bibitem{Shi}Y.Shi: ``Quantum entanglement in second-quantized
  condensed matter systems'', J.Phys. A 37(2004)6807
\bibitem{Wolf}M.C.Banuls,J.I.Cirac,M.M.Wolf: ``Entanglement in
  fermionic systems'', quant-ph/0705.1103
\bibitem{Shi2}Yu Shi: ``Entanglement in relativistic quantum field
  theory'', Phys.Rev. D 70(2004)105001
\bibitem{Haag}R.Haag: ``Local Quantum Physics'', Springer, Berlin 1992
\bibitem{Wald}R.Wald: ``Quantum Field Theory in Curver Space-Time'',
  Univ. Chicago Pr., Chicago 1994

\bibitem{Cardy1}P.Calabrese,J.Cardy: ``Entanglement Entropy and
  Quantum Field Theory: A Non-Technical Introduction'', Conference
  ``Entanglement in Physical and Information Sciences'', Pisa
  Dec. 2004, quant-ph/0505193
\bibitem{Cardy2}P.Calabrese,J.Cardy: ``Entanglement Entropy and
  Quantum Field Theory'', J.Stat.Mech. P06002 (2004), hep-th/0405152
\bibitem{Peschel}I.Peschel: ``On the Entanglement Entropy for a XY
  Spin Chain'', J.Stat.Mech. P12005 (2004), cond-mat/0410416
\bibitem{Vidal}G.Vidal,J.L.Latorre,E.Rico,A.Kitaev: ``Entanglement in
  Quantum Critical Phenomena'', Phys.Rev.Lett. 90(2003)227902-1
\bibitem{Korepin1}Heng Fan,V.Korepin,V.Roychowdhury: ``Entanglement in
  a Valence-Bond-Solid State'', Phys.Rev.Lett. 93(2004)227203, quant-ph/0406067
\bibitem{Korepin2}B.-Q.Jin,V.Korepin: ``Quantum Spin Chain, Toeplitz
  Determinants and Fisher-Hartwig Conjecture'',
  Journ.Stat.Phys. 116(2004)79, quant-ph/0304108
\bibitem{Korepin3}Heng Fan,V.Korepin,V.Roychowdhury,C.Hadley,S.Bose:
  `` Boundary Effects to Entropy and Entaglement of the Spin-1
  Valence-Bond Solid'', quant-ph/0605133
\bibitem{Verstraete1}F.Verstraete,M.A.Martin-Delgado,J.I.Cirac:
  ``Diverging Entanglement Length in Gapped Quantum Spin Systems'',
  Phys.Rev.Lett. 92(2004)087201, quant-ph/0311087
\bibitem{Verstraete2}F.Verstraete,M.Popp,J.I.Cirac: ``Entanglement
  versus Correlations in Spin Systems'',
  Phys.Rev.Lett. 92(2004)027901, quant-ph/0307009
\bibitem{White}S.R.White: ``Density matrix Formulation for Quantum
  Renormalisation Group'', Phys.Rev.Lett. 69(1992)2863
\bibitem{Nielsen}T.J.Osborne,M.A.Nielsen: ``Entanglement, Quantum
  Phase Transition, and Density Matrix Renormalisation'',
  Quant.Inf.Proc. 1(2002)45, quant-ph/0109024
\bibitem{Legeza}\"O.Legeza,J.S\'{o}lyom: ``Quantum Data Compression
  and the Density Renormalisation Group'', cond-mat/0401136
\bibitem{Farkas}S.Farkas,Z.Zimbor\'{a}s: ``On the Sharpness of the
  Zero-Entropy-Density Conjecture'', Journ.Math.Phys. 46(2005)123301, math-ph/0507022
\bibitem{Shi3}Yu Shi: ``Quantum disentanglement in long-range orders
  and spontaneous symmetry breaking'', Phys.Lett. A 309(2003)254 
\bibitem{Ruelle}D.Ruelle: ``Statistical Mechanics, Rigorous Results'',
  Benjamin Inc. N.Y. 1969

\bibitem{Reed}M.Reed,B.Simon: ``Methods of Modern Mathematical
  Physics'' vol. I, Academ.Pr., London 1980
\bibitem{Thirring}W.Thirring: ``Lehrbuch der Mathematischen Physik'',
  vol.4, ``Quantenmechanik grosser Systeme'', Springer, Wien 1980
\bibitem{Wehrl}A.Wehrl: ``General Properties of Entropy'',
  Rev.Mod.Phys. 50(1978)221
\bibitem{Plenio2}J.Eisert,C.Simon,M.B.Plenio: ``On the quantification
  of entanglement in infinite-dimensional quantum systems'', J.Phys.A
  35(2002)3911, quant-ph/0112064

\bibitem{Schmidt}E.Schmidt: Math.Annalen 63(1907)433
\bibitem{Neumann}J.von Neumann: ``Mathematische Grundlagen der
Quantenmechanik'', Springer, Berlin 1932
\bibitem{Ekert}A.Ekert,P.L.Knight: Am.J.Phys. 63(1995)415
\bibitem{Landau}L.D.Landau,E.M.Lifschitz: ``Statistische Physik'',
  Akademie-Verlag, Berlin 1966

\bibitem{Rellich}F.Rellich: ``Perturbation Theory of Eigenvalue
  Problems'', Gordon and Breach, N.Y. 1969
\bibitem{Baumgaertel}H.Baumgaertel: ``Endlich-dimensionale
  Analytische Stoerungstheorie'', Akadem. Verlagsgesellschaft, Berlin 1972
\bibitem{Achieser}N.I.Achieser,I.M.Glasmann: ``Theorie der Linearen
  Operatoren im Hilbert-Raum'', Akademie-Verlag, Berlin 1968 
\bibitem{Kato}T.Kato: ``Perturbation Theory for Linear Operators'',
  Springer, N.Y. 1966
\bibitem{Dawydow}A.Dawydow: ``Quantenmechanik'' sections 49, 50, Deutscher Verlag der
  Wissenschaften, Berlin 1974
\bibitem{Riesz}F.Riesz,B.SZ.-Nagy: ``Vorlesungen ueber
  Funktionalanalysis'', Deutscher Verlag der Wissenschaften, Berlin
  1956
\bibitem{Weinstein}A.Weinstein,W.Stenger: ``Intermediate Problems for
  Eigenvalues'', Academic Pr., N.Y.1972
\bibitem{Bhatia}R.Bhatia: ``Matrix Analysis'', Springer, Berlin 1996
\bibitem{Becker}R.Becker: ``Theorie der Waerme'', Springer, Berlin
  1966
\bibitem{Tolman}R.C.Tolman: ``The Principles of Statistical
  mechanics'', Dover, N.Y. 1979
\bibitem{Schroedinger}E.Schroedinger: ``Statistical Thermodynamics'',
  Cambridge Univ. Pr., Cambridge 1946
\bibitem{Fowler}R.H.Fowler: ``Statistical Mechanics'',
  Cambridge Univ. Pr., Cambridge 1936

}

\end{thebibliography}

\end{document}